\documentclass[aps,prb,twocolumn,preprintnumbers,amsmath,amssymb,superscriptaddress]{revtex4}%

\usepackage{graphicx}%
\usepackage{dcolumn}
\usepackage{amsmath}
\usepackage{color}

\begin{document}

\title{Multiple-gap response of type-I noncentrosymmetric BeAu superconductor}
%



\author{Rustem Khasanov}
 \email{rustem.khasanov@psi.ch}
 \affiliation{Laboratory for Muon Spin Spectroscopy, Paul Scherrer Institute, 5232 Villigen, Switzerland}

\author{Ritu Gupta}
 \affiliation{Laboratory for Muon Spin Spectroscopy, Paul Scherrer Institute, 5232 Villigen, Switzerland}

\author{Debarchan Das}
 \affiliation{Laboratory for Muon Spin Spectroscopy, Paul Scherrer Institute, 5232 Villigen, Switzerland}


\author{Alfred Amon}
 \affiliation{Max-Planck-Institut f\"{u}r Chemische Physik fester Stoffe, N\"{o}thnitzer Stra{\ss}e 40, 01187 Dresden, Germany}

\author{Andreas Leithe-Jasper}
 \affiliation{Max-Planck-Institut f\"{u}r Chemische Physik fester Stoffe, N\"{o}thnitzer Stra{\ss}e 40, 01187 Dresden, Germany}




\author{Eteri Svanidze}
 \affiliation{Max-Planck-Institut f\"{u}r Chemische Physik fester Stoffe, N\"{o}thnitzer Stra{\ss}e 40, 01187 Dresden, Germany}

\begin{abstract}
Precise measurements of the thermodynamic critical field ($B_{\rm c}$) in type-I noncentrosymmetric superconductor BeAu were performed by means of the muon-spin rotation/relaxation technique. The temperature evolution of $B_{\rm c}$ can not be described within the single gap scenario and it requires the presence of at least two different types of the superconducting order parameters. The self-consistent two-gap approach, adapted for analysis of $B_{\rm c}(T)$ behavior, suggests the presence of two superconducing energy gaps with the gap to $T_{\rm c}$ ratios $2\Delta/k_{\rm B}T_{\rm c}\simeq4.52$ and $\simeq2.37$ for the big and the small gap, respectively. This implies that the superconductivity in BeAu is unconventional and that the supercarrier pairing occurs at various energy bands.
\end{abstract}

\maketitle

BeAu is an old known superconductor with the transition temperature $T_{\rm c}\simeq 3.2$~K. Superconductivity in BeAu was originally discovered by Matthias in 1959,\cite{Matthias_JPCS_1959} {\it i.e.} just in two years after the formulation of the BCS theory.\cite{BCS_PR_1957}  In this short report, Matthias was noted the absence of a superconductivity in a pure Be and Au (Be was later found to have $T_{\rm c}\simeq26$~mK, Ref.~\onlinecite{Poole_Book_2014}) and performed a search within the gold-rich site of the Be-Au phase diagram. The superconductivity was found to appear in a stoichiometric ({\it i.e.} 1:1 Be to Au ratio) BeAu sample.\cite{Matthias_JPCS_1959}

Recently, the interest to BeAu was renewed.\cite{Rebar_PhD-Thesis_2015, Amon_PRB_2018, Rebar_PRB_2019, Singh_PRB_2019, Beare_PRB_2019} This mostly relates to the realisation of their non-centrosymmetric crystal structure, which was expected to give rise to unconventional superconductivity due to spin-orbit coupling and/or mixed singlet/triplet pairing state (see {\it e.g.} Refs.~\onlinecite{Bauer_PRL_2004, Kaur_PRL_2005, Yuan_PRL_2006, Khasanov_PRB_2006, Schnyder_PRB_2008, Maisuradze_PRL_2009, Hafliger_JSNM_2009, Maisuradze_PRB_2010} and references therein).
In addition, the $B20$ FeSi-type of the crystal structure of BeAu becomes particualry interesting since such materials were predicted to host chiral fermions in topological semimetals.\cite{Bradlyn_Science_2016, Tang_PRL_2017, Chang_PRL_2017}  Moreover, $B20$ structure is the only known crystal structure
for bulk magnetic skyrmions in materials such as MnSi, Fe$_{1-x}$Co$_x$Si, FeGe, MnGe,  Cu$_2$OSeO$_3$ {\it etc}.\cite{Tonomura_NanoLett_2012, Yu_Naure_2010, Kanazawa_NatMat_2011, Kanazawa_PRB_2012, Seci_Science_2012} All these make BeAu an intriguing candidate material to search for unconventional superconductivity, associated with its noncentrosymmetric crystal structure in combination with the possible existence of exotic quasiparticles.

The previously published papers agreed that BeAu is characterised by an isotropic superconducting energy gap and, predominantly, by an $s-$wave spin-singlet pairing in the weak coupling regime.\cite{Rebar_PhD-Thesis_2015, Amon_PRB_2018, Rebar_PRB_2019, Singh_PRB_2019, Beare_PRB_2019} There is, however, disagreement on the type of superconductivity. An intensive characterization of BeAu by means of the specific heat, dc-magnetization, ac-susceptibility, and resistivity, performed by Rebar {\it et al.}\cite{Rebar_PhD-Thesis_2015, Rebar_PRB_2019} reveal a crossover from type-I to type-II superconductivity at approximately 1.2~K, thus putting BeAu to the class of the so-called type-1.5 or type-II/I superconductors.\cite{Krageloh_PLA_1969, Auer_PRA_1973}  On the other hand, purely type-I behaviour was reported by Singh {\it et al.}\cite{Singh_PRB_2019} and Beare {\it et al.}\cite{Beare_PRB_2019} based on the results of dc-magnetization, specific heat and muon-spin rotation/relaxation experiments. In addition, the results of De Haas -  van Alphen experiments combined with the Density Functional theory (DFT) band structure calculations reveal the presence of multiple (at least three) conductive bands crossing the Fermi level.\cite{Rebar_PhD-Thesis_2015, Rebar_PRB_2019} Consequently, if superconductivity in BeAu sets in different energy bands, effects of a multiple-band structure are expected to influence the supercarrier formation. This is, {\it e.g.}, the case for the famous two-gap superconductor MgB$_2$, the broad variety of Fe-based and cuprate high-temperature superconductors {\it etc.} (see {\it e.g.} Refs.~\onlinecite{Bouquet_PRL_2001, Tsuda_PRL_2001, Chen_PRL_2001, Szabo_PRL_2001, Gonelli_PRL_2002, Khasanov_La214_PRL_2007, Khasanov_Y123_PRL_2007, Khasanov_Y123_JSNM_2008, Ideta_PRL_2010, Kunisada_PRL_2017, Ding_EPL_2008, Evtushinsky_NJP_2009, Khasanov_Ba122_PRL_2009, Khasanov_Sr122_PRL_2009, Khasanov_SrPtP_PRB_2014, Khasanov_1144_PRB_2019} and references therein). Note however, that until now the multiple-band features were not detected on the measured thermodynamical quantities and the superconductivity of BeAu was treated within the single-band approach.\cite{Rebar_PhD-Thesis_2015,  Amon_PRB_2018, Rebar_PRB_2019, Singh_PRB_2019, Beare_PRB_2019}

In this paper we report on the results of the precise measurements of the thermodynamical critical field $B_{\rm c}$ of BeAu by means of the muon-spin rotation/relaxation ($\mu$SR) technique. The analysis of $B_{\rm c}(T)$ within the self-consistent two-gap approach reveals the presence of two superconducing energy gaps with the ratio $2\Delta_1/k_{\rm B}T_{\rm c}\simeq4.52$ and $2\Delta_2/k_{\rm B}T_{\rm c}\simeq2.37$ for the big ($\Delta_1$) and the small ($\Delta_2$) gap, respectively. Our results imply that the multiple-band superconductivity can also be realized in type-I superconducting materials, in analogy with that reported in Refs.~\onlinecite{Ruby_PRL_2015,Singh_PRB_2018}.

\begin{figure*}[htb]
\includegraphics[width=1.0\linewidth]{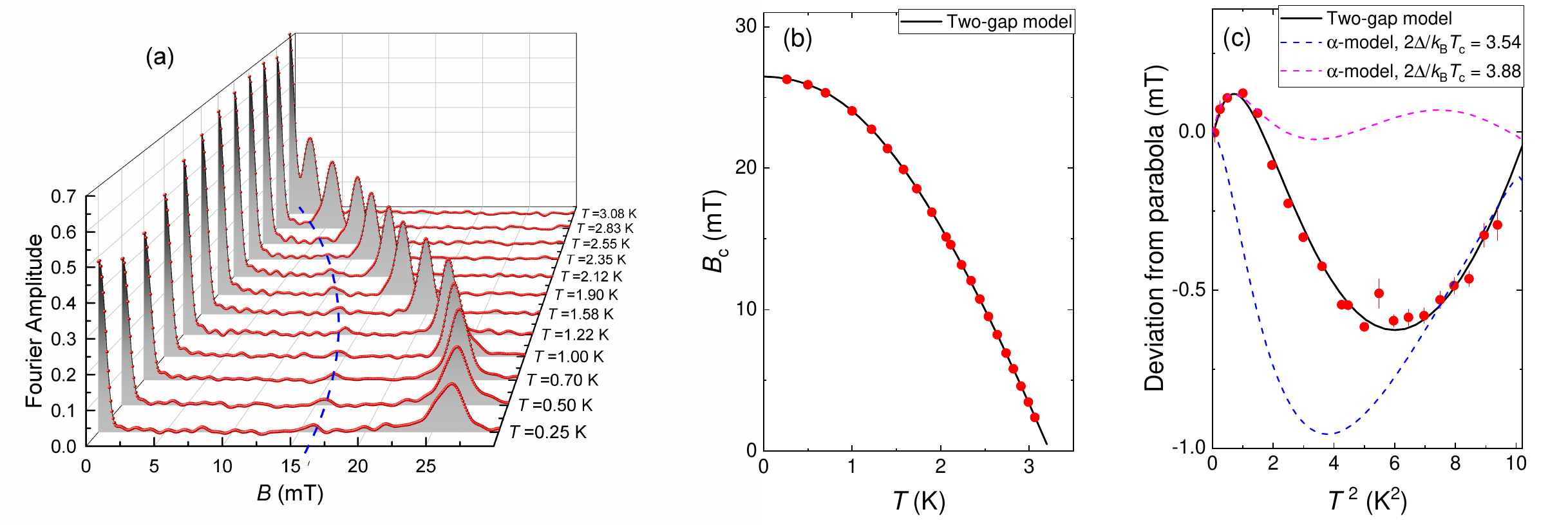}
%
\caption{(a) Fast Fourier transform of the TF-$\mu$SR time spectra of BeAu for several $T$ and $B_{\rm ex}$ values. Dashed blue line represents the position of the external  field $B_{\rm ex}$. Temperatures are marked on the $y$-axis.  The tiny background peak at $B=B_{\rm ex}$ position corresponds to the muons which missed the sample. (b) The temperature evolution of the thermodynamical critical field of BeAu obtained from the $B>B_{\rm ex}$ peak position. The solid line is the fit of the self-consistent two-gap model  (Eq.~\ref{eq:final-fit}) to the experimental $B_{\rm c}(T)$. (c) The deviation of $B_{\rm c}(T)$ from parabolic dependence. The dashed blue and pink lines are expectations for the single s-wave gap. The solid black line is the self-consistent two-gap fit. See text for details. }
 \label{fig:Fourier-transform_Bc_Deviations}
\end{figure*}

The polycrystalline BeAu samples were the same as used by Beare {\it et al.} in Ref.~\onlinecite{Beare_PRB_2019}. 
The muon-spin rotation/relaxation measurements were performed at the $\pi$E1 beamline by using Dolly spectrometer (Paul Scherrer Institute, PSI Villigen, Switzerland). The $^4$He cryostat equipped with the $^3$He inset was used.  Samples were sandwiched between two pieces of a thin copper foil ($\simeq 10$~$\mu$m each), which are transparent to positive surface muons used in our studies. The foil/sample assembly was screwed to the frame-shaped sample holder and attached, via $\simeq 1$~cm in diameter thick copper rod, to the bottom part of the $^3$He inset. Such construction allows, on the one side, to ensure on a good thermal contact between the sample and the $^3$He pot and, on the other side, permits the use of a so-called 'Veto' mode of the $\mu$SR spectrometer. The 'Veto' mode rejects the muons missing the sample and, as a consequence, reduces the background of the $\mu$SR response to almost zero [see tiny background peaks following the dashed blue line in Fig.~\ref{fig:Fourier-transform_Bc_Deviations}~(a) and Ref.~\onlinecite{Amato_GPS_2017} for the detailed explanations of the 'Veto' mode principle]. The external magnetic field $B_{\rm ex}$ was applied parallel to the direction of the muon momentum and perpendicular to the initial muon spin direction, which corresponds to the transverse field (TF) $\mu$SR geometry. Several elliptically shaped discs of BeAu  (0.5 mm thick and roughly 3.0x4.0~mm$^2$ in size), were used. The field was applied perpendicular to the flat faces of the samples. The experiments were performed in the temperature range of 0.25–5~K and in the field range of 0.5 to 50 mT.

The TF-$\mu$SR experiments were performed in the intermediate state of BeAu superconducting sample, {\it i.e.} when the sample volume is separated on the normal state and the superconducting (Meissner) state domains.\cite{Tinkham_book_1975, Poole_Book_2014, Singh_PRB_2019, Beare_PRB_2019, deGennes_Book_1966, Kittel_Book_1996, Prozorov_PRL_2007, Prozorov_NatPhys_2008, Khasanov_Bi-II_PRB_2019, Karl_PRB_2019, Khasanov_Ga-II_arxiv} The $B-T$-scan measuring scheme, as discussed by Karl {\it et al.} in Ref.~\onlinecite{Karl_PRB_2019}, was used. Each measured point was reached in two steps: first -- by stabilizing the temperature and second -- by swiping the field to $B>B_{\rm c}$ (up to 30~mT in our case) and decreasing it back to the measuring one. The idea of such a scheme is two-fold: (i) to keep unchanged the volume parts of the sample occupied by the normal state ($f_{\rm N}$) and the superconducting state ($f_{\rm S}$) domains ($f$ denotes the volume fraction) and (ii) to maintain similarly distributed domain patterns for each $B-T$ measuring point. Note that the shape of the domain patterns in type-I supercondcutors depend strongly on the magnetic history.\cite{Prozorov_PRL_2007, Prozorov_NatPhys_2008}  The $B-T$ points were taken along the $\simeq 0.6 \cdot B_{\rm c}(T)$ line by considering the $B_{\rm c}(T)$ curve as is determined in Refs.~\onlinecite{Singh_PRB_2019, Beare_PRB_2019} (see the Supplemental part, Ref.~\onlinecite{Supplemental_part}, for details).

The magnetic field distributions in the BeAu sample, as obtained from the Fourier transform of TF-$\mu$SR time spectra, are shown in Fig.~\ref{fig:Fourier-transform_Bc_Deviations}~(a). The blue dashed line represents the position of the external field $B_{\rm ex}$. Down to the lowest temperature studied ($T\simeq 0.25$~K), the signal  splits on two peaks positioned at $B=0$ and $B>B_{\rm ex}$, respectively. Such distributions are typical for type-I superconducting materials in the intermediate state.\cite{Khasanov_Bi-II_PRB_2019, Karl_PRB_2019, Khasanov_Ga-II_arxiv} Our results, therefore, confirm conclusion of Refs.~\onlinecite{Singh_PRB_2019, Beare_PRB_2019} on type-I superconductivity of BeAu.  The area below the peaks corresponds to the superconducting (Meissner) state ($f_{\rm S}$) and the normal state ($f_{\rm N}$) volume fractions, while the position of  $B>B_{\rm ex}$ peak determines the value of the thermodynamic critical field $B_{\rm c}$.\cite{Khasanov_Bi-II_PRB_2019, Karl_PRB_2019, Khasanov_Ga-II_arxiv} The temperature evolution of $B_{\rm c}$, obtained from the fit of TF-$\mu$SR data, is shown in Fig.~\ref{fig:Fourier-transform_Bc_Deviations}~(b). The details of the data analysis procedure, as well as the temperature dependencies of the fitting parameters, are presented in the Supplemental part of the manuscript.\cite{Supplemental_part}

The deviation of the $B_{\rm c}$ {\it vs.} $T$ curve from the parabolic function: $D(T^2)=B_{\rm c}(T^2)-B_{\rm c}(0)[1-(T/T_{\rm c})^2]$ is shown in Fig.~\ref{fig:Fourier-transform_Bc_Deviations}~(c). Following Refs.~\onlinecite{Padamsee_JLTP_1973, Johnston_SST_2013} the shape of $D(T^2)$ function depends strongly on the $2\Delta/k_{\rm B}T_{\rm c}$ ratio and it is also expected to be  sensitive to the symmetry of the superconducting energy gap.

Bearing in mind the isotropic single gap behavior reported in Refs.~\onlinecite{Rebar_PhD-Thesis_2015,Rebar_PRB_2019}, the initial analysis of $B_{\rm c}(T)$ dependence of BeAu was performed by means of the phenomenological $\alpha$-model within the single $s-$wave gap approach.\cite{Padamsee_JLTP_1973, Johnston_SST_2013}
The fit procedure and the results obtained from such analysis are described in the Supplemental part.\cite{Supplemental_part} The analysis reveals that the $D(T^2)$ data {\it can not} be described within the single-gap scenario. The low- and the high-temperature parts of $D(T^2)$ require rather different $2\Delta/k_{\rm B}T_{\rm c}$ values. To put this statement into evidence, the dashed pink and blue lines in Fig.~\ref{fig:Fourier-transform_Bc_Deviations}~(c) correspond to the theoretical  $D(T^2)$ curves with $2\Delta/k_{\rm B} T_{\rm c}=3.88$ and $3.54$, respectively. The disagreement between the single $s$-wave gap analysis and the experimental data implies, therefore, that in BeAu a more complicated gap scenario needs to be considered.

By taking into account the presence of multiple bands crossing the Fermi level, as is reported in de Haas - van Alphen experiments,\cite{Rebar_PhD-Thesis_2015, Rebar_PRB_2019} as well as confirmed by DFT calculations,\cite{Amon_PRB_2018,Rebar_PhD-Thesis_2015, Rebar_PRB_2019} the further analysis of the temperature evolution of the thermodynamic critical field was performed within the self-consistent two $s$-wave gap scenario.

Following Refs.~\onlinecite{Bussmann-Holder_EPB_2004, Bussmann-Holder_Arxiv_2009, Khasanov_PRL_2010}, within the two-gap approach the coupled $s$-wave gap equations are described as:
\begin{widetext}
\begin{equation}
\Delta_1 (T)= \int_{0}^{\omega_{D}}\frac{n_1V_{11}\Delta_1(T)}{\sqrt{E^2+\Delta_1^2(T)}}\tanh \frac{\sqrt{E^2+\Delta_1^2(T)}}{2k_BT}dE + \int_{0}^{\omega_{D}}\frac{n_2V_{12}\Delta_2(T)}{\sqrt{E^2+\Delta_2^2(T)}}\tanh \frac{\sqrt{E^2+\Delta_2^2(T)}}{2k_BT}dE, \nonumber
\end{equation}
\begin{equation}
\Delta_2 (T) = \int_{0}^{\omega_{D}}\frac{n_1V_{21}\Delta_1(T)}{\sqrt{E^2+\Delta_1^2(T)}}\tanh \frac{\sqrt{E^2+\Delta_1^2(T)}}{2k_BT}dE +
\int_{0}^{\omega_{D}}\frac{n_2V_{22}\Delta_2(T)}{\sqrt{E^2+\Delta_2^2(T)}}\tanh \frac{\sqrt{E^2+\Delta_2^2(T)}}{2k_BT}dE.
 \label{eq:Coupled-Gaps_Full}
\end{equation}
\end{widetext}
Here $\Delta_1(T)$ and $\Delta_2(T)$ are temperature evolutions of gaps within the band 1 and 2; $n_1$ and $n_2$ are the partial density of states for each band at the Fermi level ($n_1+n_2=1$); $V_{11}$ ($V_{22}$) and $V_{12}$ ($V_{21}$) are the intraband and the interband interaction potentials, respectively. For simplicity, it is also assumed that the Debye frequency ($\omega_{\rm D}$) is the same for both bands.

\begin{figure}[htb]
\includegraphics[width=0.85\linewidth]{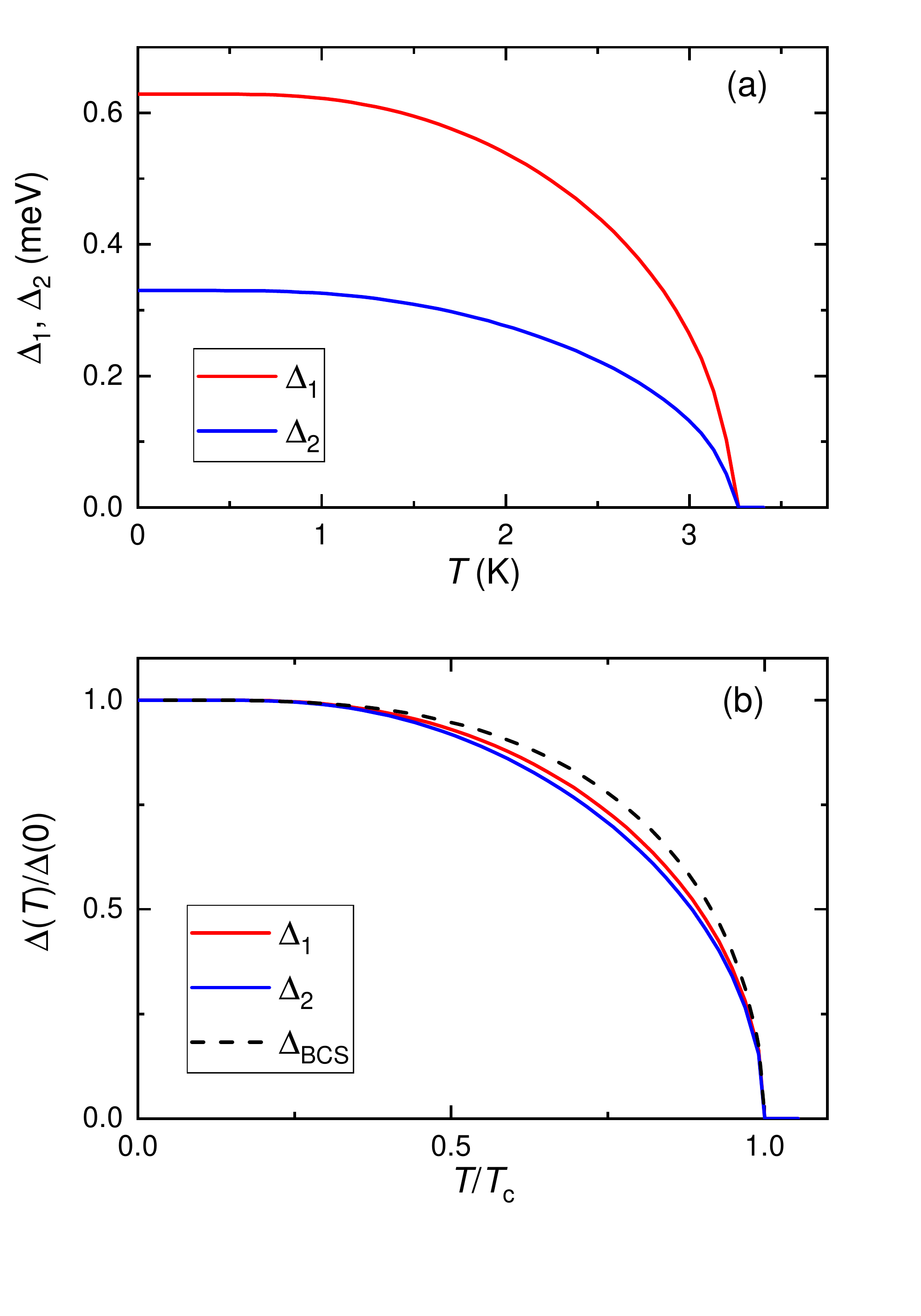}
%
\caption{(a) Temperature evolutions of the big ($\Delta_1$) and the small ($\Delta_2$) superconducting gaps obtained within the framwork of the self-consistent two-gap model. (b) Comparison of temperature dependencies of $\Delta_1$ and $\Delta_2$ with the weak-coupled BCS prediction.\cite{Muehlschlegel_ZPhys_1959} }
 \label{fig:Gaps}
\end{figure}

The thermodynamic critical field $B_{\rm c}$ was determined from the normal state ($F_{\rm N}$) and the superconducting state ($F_{\rm S}$) free energy difference:\cite{Tinkham_book_1975}
\begin{equation}
\frac{B_{\rm c}^2(T)}{8 \pi}= F_{N}(T) - F_{\rm S}(T).
 \label{eq:Bc_FreeEn}
\end{equation}
Considering the presence of two superconducting energy bands, the second term in the right-hand site of the above equation  consists of two contibutions:
\begin{equation}
F_{\rm S}(T) = n_1 F_{\rm S}[T,\Delta_1(T)]+n_2 F_{\rm S}[T,\Delta_2(T)]
 \label{eq:F_S}
\end{equation}
Following Johnston,\cite{Johnston_SST_2013} the individual  free energy components are further derived as:
\begin{equation}
{F_{\rm N}(T)}=-\frac{\gamma_{\rm e } T^2}{2}
 \label{eq:FreeEn_Ns}
\end{equation}
and
\begin{eqnarray}
F_{\rm S}[T,\Delta(T)]  & = &-\frac{3 \gamma_e }{4 \pi^2 k_{\rm B}^2}
\bigg[  \Delta(T)^2 + \bigg. \nonumber \\
&& \left. 4 \int_0^\infty f(E,T) \frac{2\varepsilon^2 + \Delta(T)^2}{E} d\varepsilon \right].  \label{eq:FreeEn_Ss}
\end{eqnarray}
Here, $f(E,T)=[\exp( E/k_{\rm B}T) + 1]^{-1}$ is the Fermi function, $E=E[\varepsilon,\Delta (T)]=\sqrt{ \varepsilon^2 + \Delta(T)^2 }$ is the quasiparticle energy, and $\gamma_{\rm e}$ is the normal state electronic specific heat coefficient. Note that the Eqs.~\ref{eq:Bc_FreeEn},  \ref{eq:FreeEn_Ns}, and  \ref{eq:FreeEn_Ss} are expressed in cgs units.\cite{Johnston_SST_2013}
Finally, the equation used to fit to the experimental $B_{\rm c}(T)$ data presented in Fig.~\ref{fig:Fourier-transform_Bc_Deviations}~(b) took the form:
\begin{equation}
\frac{B_{\rm c}^2(T)}{B_{\rm c}^2(0)}= \frac{F_{N}(T) - n_1 F_{\rm S}[T,\Delta_1(T)] -n_2 F_{\rm S}[T,\Delta_2(T)]}
{F_{N}(0) - n_1 F_{\rm S}(0,\Delta_1) -n_2 F_{\rm S}(0,\Delta_2)}.
  \label{eq:final-fit}
\end{equation}
Here, $\Delta_1=\Delta_1(T=0)$ and $\Delta_2=\Delta_2(T=0)$ are the zero-temperature values of the superconducting enery gap.
During the fit the following parameters were varied: $B_{\rm c}(0)$, $n_1$, $V_{11}$, $V_{22}$, $V_{12}$, and $V_{21}$.

The fit requires solving two coupled nonlinear equation (Eq.~\ref{eq:Coupled-Gaps_Full}). We used Mathcad 15.0 with the Levenberg-Marquardt non-linear equation solver.\cite{MathCad} The results of the analysis, with the Debye frequency $\omega_{\rm D} \simeq25.4$~meV,\cite{Rebar_PRB_2019} are shown by black solid lines in panels (b) and (c) of Fig.~\ref{fig:Fourier-transform_Bc_Deviations}. Obviously, the self-consistent two-gap model describes the experimental data remarkably well. The fit parameters are: $B_{\rm c}(0)\simeq 26.04$~mT, $n_1\simeq 0.65$, $V_{11}\simeq 0.246$, $V_{22}\simeq 0.343$, $V_{12}\simeq 0.321$, and $V_{21}\simeq 0.0727$. The value of the superconducting transition temperature, corresponding to $B_{\rm c}= 0$, was found  $T_{\rm c}\simeq 3.232$~K. The temperature evolutions of the big and the small gap and the comparison of their temperature dependencies with the weak-coupled BCS prediction,\cite{Muehlschlegel_ZPhys_1959} are presented in Fig.~\ref{fig:Gaps}.

From the analysis of $B_{\rm c}(T)$ of BeAu within the two-gap approach, the following important points emerge:\\
 (i) The partial density of states for the band(s), where the big superconducting energy gap opens, is almost twice as high as that for the band(s) with the small gap: $n_1\simeq 2 \cdot n_2$. The results of the band structure calculation, presented in Ref.~\onlinecite{Amon_PRB_2018}, reveal that the partial density of states at the Fermi level is dominated by Be 2$p$ , Au 5$p$, and Au 6$p$ bands, so that $n_{\rm Au5{\it p}}\simeq n_{\rm Au6{\it p}}\simeq 0.6-0.7n_{\rm Be2{\it p}}$. This suggests that the big gap may be opened in Be 2$p$ and one of Au band ($5p$ or $6p$), while the other Au band ($6p$ or $5p$) carries the small superconducting gap. \\
 (ii) The zero-temperature values of the big and the small gaps were estimated to be $\Delta_1\simeq 0.62$~meV
 and $\Delta_2\simeq 0.32$~meV. The corresponding gap to $T_{\rm c}$ ratios are: $2\Delta_1/k_{\rm B}T_{\rm c}\simeq4.52$ and $2\Delta_2/k_{\rm B}T_{\rm c}\simeq2.37$. Such values suggests strong-coupling within the first and weak-coupling within the second band, respectively. \\
 (iii) The interband coupling potentials $V_{12}\simeq 0.321$ and $V_{21}\simeq 0.0727$ have relatively high values. In particular, $V_{12}$ is comparable to $V_{22}$ and is even higher than  $V_{11}$. This suggests strong coupling between the bands, which also results in an almost identical temperature dependencies of the big [$\Delta_{1}(T)$] and the small [$\Delta_{2}(T)$] gap, see Fig.~\ref{fig:Gaps}~(b). Note that in the case of $V_{12},V_{21}\ll V_{11},V_{22}$, $\Delta_{1}(T)/\Delta_1$ and $\Delta_{2}(T)/\Delta_2$ are very much different.\cite{Kogan_PRB_2009, Khasanov_PRL_2010, Gupta_arxiv_2019} \\
 (iv) The temperature dependencies of both gaps, presented in Fig.~\ref{fig:Gaps}~(b), are {\it weaker} than the weak-coupled BCS prediction.
 We believe that such effects are related to the BeAu system rather than to the model itself. The lack of absence of the experimental data for BeAu does not allow to make a comparison of $\Delta(T)$'s dependencies obtained in the present study with the directly measured gap values. Note however, that the recent work of Kim {\it et al.}\cite{Kim_Symmetry_2019} on MgB$_2$, where such data become accessible, reveal that $\Delta(T)$'s obtained within the framework of the self-consistent model match rather well temperature evolutions of superconducting energy gaps measured in tunneling experiments. More interesting is that the 'weakening' of the gaps in MgB$_2$ is also observed both experimentally and theoretically.

It is worth to note here that due to the noncentrosymmetric crystal structure of BeAu, the phase of the superconducting order parameter may change between the different energy bands. In relation to the two $s$-wave gap behaviour reported here, both $s^{++}$ and $s^{+-}$ scenarios could be realised in BeAu. The above presented self-consistent model is, however, 'phase unsensitive' and does not allow to distinguish between them. An additional check could be performed by studying the influence of nonmagnetic impurities on the supercondcuting transition temperature. As shown in Refs.~\onlinecite{Bang_PRB_2009} and \onlinecite{Trevisan_PRB_2018}, the transition temperature $T_{\rm c}$ in $s^{+-}$ superconductors is fast suppressed due to the pair-breaking promoted
by interband impurity scattering, while it remains only marginally affected for $s^{++}$ materials. A series of $T_{\rm c}$ and $B_c(T)$ measurements of BeAu samples with a different amount of nonmagnetic impurities could be highly reliable and may help to clarify this question.

To summarise, the precise measurements of the thermodynamic critical field $B_{\rm c}$ in noncentrosymmetric BeAu superconductor were performed by means of the muon-spin rotation/relaxation technique. The main results are the following: \\
 (i) Within the full temperature range studied, the superconductivity in BeAu remains of a type-I. No indications of type-II superconductivity for $T\lesssim 1.2$~K and at least down to $T \simeq 0.25$~K was detected. \\
 (ii) The value of the thermodynamic critical field at $T=0$ and the transition temperature $T_{\rm c}$ were found to be $B_{\rm c}(0)\simeq26.04$~mT and $T_{\rm c}\simeq 3.232$~K, in agreement with the previously published data.\cite{Singh_PRB_2019, Beare_PRB_2019} \\
 (iii) The single $s$-wave gap approach does not allow to describe the temperature evolution of the thermodynamic critical field $B_{\rm c}$, thus calling for the presence of an additional supercondcuting energy gap(s). \\
 (iv) The self-consistent two-gap model, adapted for analysis of $B_{\rm c}(T)$ behavior, was developed. It takes into account the intraband coupling within the each individual band, as well as the interband coupling between the bands. The model was found to describe accurately the experimental $B_{\rm c}(T)$ data of BeAu.\\
 (v) The gap to $T_{\rm c}$ ratios were estimated to be $2\Delta_1/k_{\rm B}T_{\rm c}\simeq4.52$ and $2\Delta_2/k_{\rm B}T_{\rm c}\simeq2.37$. This implies that the strong- and the weak-coupling occur within the first and the second band, respectively.\\
To conclude, our results suggest that the superconductivity in BeAu is unconventional and that the supercarrier pairing takes place in different energy bands. More important, our results imply that the multiple-gap behaviour is also realized {\it in type-I} superconducting materials.

This work was performed at the Swiss Muon Source (S$\mu$S), Paul Scherrer Institute (PSI, Switzerland). The work of RG is supported
by the Swiss National Science Foundation (SNF-Grant No. 200021-175935). The authors acknowledge the technical support of Toni Shiroka and Chris Baines. Juri Grin is acknowledged for his steady support.
RK acknowledges helpful discussions with Zurab Guguchia.


\begin{thebibliography}{99}
%
\bibitem{Matthias_JPCS_1959} B.T. Matthias, J. Phys. Chem. Solids {\bf 10}, 342 (1959).

\bibitem{BCS_PR_1957} J. Bardeen, L. N. Cooper, and J. R. Schrieffer Phys. Rev. {\bf 106}, 162 (1957).

\bibitem{Poole_Book_2014} C. Poole, H. Farach, R. Creswick, and  R. Prozorov, {\it Superconductivity 3rd Edition} (Elseiver: Amsterdam,
    2014).

\bibitem{Rebar_PhD-Thesis_2015} Drew Rebar, Exploring superconductivity in chiral structured BeAu, Ph.D. dissertation, Louisiana State University, 2015.

\bibitem{Amon_PRB_2018} A. Amon, E. Svanidze, R. Cardoso-Gil, M. N. Wilson, H. Rosner, M. Bobnar, W. Schnelle, J. W. Lynn, R. Gumeniuk, C. Hennig, G. M. Luke, H. Borrmann, A. Leithe-Jasper, and Yu. Grin, Phys. Rev. B {\bf 97}, 014501 (2018).

\bibitem{Rebar_PRB_2019} D. J. Rebar, S. M. Birnbaum, J. Singleton, M. Khan, J. C. Ball, P. W. Adams, J. Y. Chan, D. P. Young, D. A. Browne, and J. F. Di Tusa, Phys. Rev. B {\bf 99}, 094517 (2019).

\bibitem{Singh_PRB_2019} D. Singh, A. D. Hillier, and R. P. Singh, Phys. Rev. B {\bf 99}, 134509 (2019).

\bibitem{Beare_PRB_2019} J. Beare, M. Nugent, M. N. Wilson, Y. Cai, T. J. S. Munsie, A. Amon, A. Leithe-Jasper, Z. Gong, S. L. Guo, Z. Guguchia, Y. Grin, Y. J. Uemura, E. Svanidze, and G. M. Luke Phys. Rev. B {\bf 99}, 134510 (2019).

\bibitem{Bauer_PRL_2004} E. Bauer, G. Hilscher, H. Michor, Ch. Paul, E. W. Scheidt, A. Gribanov, Yu. Seropegin, H. N\"{o}el, M. Sigrist, and P. Rogl, Phy. Rev. Lett. {\bf 92}, 027003 (2004).

\bibitem{Kaur_PRL_2005} R. P. Kaur, D. F. Agterberg, and M. Sigrist, Phys. Rev. Lett. {\bf 94}, 137002 (2005).

\bibitem{Yuan_PRL_2006} H. Q. Yuan, D. F. Agterberg, N. Hayashi, P. Badica, D. Vandervelde, K. Togano, M. Sigrist, and M. B. Salamon, Phys. Rev. Lett. {\bf 97}, 017006 (2006).

\bibitem{Khasanov_PRB_2006} R. Khasanov, I. L. Landau, C. Baines, F. La Mattina, A. Maisuradze, K. Togano, and H. Keller, Phys. Rev. B {\bf 73}, 214528 (2006).

\bibitem{Schnyder_PRB_2008} A. P. Schnyder, S. Ryu, A. Furusaki, and A. W. W. Ludwig, Phys. Rev. B {\bf 78}, 195125 (2008).

\bibitem{Maisuradze_PRL_2009} A. Maisuradze, M. Nicklas, R. Gumeniuk, C. Baines, W. Schnelle, H. Rosner, A. Leithe-Jasper, Yu. Grin, and R. Khasanov, Phys. Rev. Lett. 103, 147002 (2009).

\bibitem{Hafliger_JSNM_2009}  P.S. H\"{a}fliger,  R. Khasanov,  R. Lortz,  A. Petrovi\'{c}, K. Togano, C. Baines, B. Graneli, and  H. Keller, J. Supercond. Nov. Magn.  {\bf 22} 337 (2009).

\bibitem{Maisuradze_PRB_2010} A. Maisuradze, W. Schnelle, R. Khasanov, R. Gumeniuk, M. Nicklas, H. Rosner, A. Leithe-Jasper, Yu. Grin, A. Amato, and P. Thalmeier, Phys. Rev. B {\bf 82}, 024524 (2010).

\bibitem{Bradlyn_Science_2016} B. Bradlyn, J. Cano, Z. Wang, M. G. Vergniory, C. Felser, R. J. Cava, and A. Bernevig, Science {\bf 353}, aaf5037 (2016).

\bibitem{Tang_PRL_2017} P. Tang, Q. Zhou, and S.-C. Zhang, Phys. Rev. Lett. {\bf 119}, 206402 (2017).

\bibitem{Chang_PRL_2017} G. Chang, S.-Y. Xu, B. J. Wieder, D. S. Sanchez, S.-M. Huang, I. Belopolski, T.-R. Chang, S. Zhang, A. Bansil, H. Lin, and M. Z. Hasan, Phys. Rev. Lett. {\bf 119}, 206401 (2017).

\bibitem{Tonomura_NanoLett_2012} A. Tonomura, Y. Xiuzhen, K. Yanagisawa, T. Matsuda, Y. Onose, N. Kanazawa, H. Park and Y. Tokura, Nano Lett. {\bf 12}, 1673 (2012).

\bibitem{Yu_Naure_2010} X. Yu, Y. Onose, N. Kanazawa, J. Park, J. Han, Y. Matsui, N. Nagaosa and Y. Tokura, Nature {\bf 465}, 901 (2010).

\bibitem{Kanazawa_NatMat_2011} X. Yu, N. Kanazawa, Y. Onose, K. Kimoto, W. Zhang, S. Ishiwata, Y. Matsui and Y. Tokura, Nature  Mater. {\bf 10}, 106, (2011).

\bibitem{Kanazawa_PRB_2012} N. Kanazawa, J. Kim, D. Inosov, J.S. White, N. Egetenmeyer, J.L. Gavilano, S. Ishiwata, Y. Onose, T. Arima, B. Keimer and Y. Tokura, Phys. Rev. B {\bf 86}, 134425 (2012).

\bibitem{Seci_Science_2012} S. Seki, X. Yu, S. Ishiwata and Y. Tokura, Science {\bf 336}, 198 (2012).

\bibitem{Krageloh_PLA_1969} U. Kr\"{a}geloh, Phys. Lett. A {\bf 28}, 657 (1969).

\bibitem{Auer_PRA_1973} J. Auer and H. Ullmaier, Phys. Rev. B {\bf 7}, 136 (1973).

\bibitem{Bouquet_PRL_2001} F. Bouquet, R. A. Fisher, N. E. Phillips, D. G. Hinks, and J. D. Jorgensen, Phys. Rev. Lett. {\bf 87}, 047001 (2001).

\bibitem{Tsuda_PRL_2001} S. Tsuda, T. Yokoya, T. Kiss, Y. Takano, K. Togano, H. Kito, H. Ihara, and S. Shin, Phys. Rev. Lett. {\bf 87}, 177006 (2001).

\bibitem{Chen_PRL_2001} X. K. Chen, M. J. Konstantinović, J. C. Irwin, D. D. Lawrie, and J. P. Franck, Phys. Rev. Lett. {\bf 87}, 157002 (2001).

\bibitem{Szabo_PRL_2001} P. Szab\'{o}, P. Samuely, J. Ka\v{c}mar\v{c}ík, T. Klein, J. Marcus, D. Fruchart, S. Miraglia, C. Marcenat, and A. G. M. Jansen, Phys. Rev. Lett. {\bf 87}, 137005 (2001).

\bibitem{Gonelli_PRL_2002} R. S. Gonnelli, D. Daghero, G. A. Ummarino, V. A. Stepanov, J. Jun, S. M. Kazakov, and J. Karpinski, Phys. Rev. Lett. {\bf 89}, 247004 (2002).

\bibitem{Khasanov_La214_PRL_2007} R. Khasanov, A. Shengelaya, A. Maisuradze, F. La Mattina, A. Bussmann-Holder, H. Keller, and K. A. M\"{u}ller, Phys. Rev. Lett. {\bf 98}, 057007 (2007)

\bibitem{Khasanov_Y123_PRL_2007} R. Khasanov, S. Str\"{a}ssle, D. Di Castro, T. Masui, S. Miyasaka, S. Tajima, A. Bussmann-Holder, and H. Keller, Phys. Rev. Lett. {\bf 99}, 237601 (2007).

\bibitem{Khasanov_Y123_JSNM_2008} R. Khasanov, A. Shengelaya, J. Karpinski, A. Bussmann-Holder, H. Keller, and K. A. M\"{u}ller, J. Supercond. Nov. Magn. {\bf 21}, 81 (2008).

\bibitem{Ideta_PRL_2010} S. Ideta, K. Takashima, M. Hashimoto, T. Yoshida, A. Fujimori, H. Anzai, T. Fujita, Y. Nakashima, A. Ino, M. Arita, H. Namatame, M. Taniguchi, K. Ono, M. Kubota, D. H. Lu, Z.-X. Shen, K. M. Kojima, and S. Uchida, Phys. Rev. Lett. {\bf 104}, 227001 (2010).

\bibitem{Kunisada_PRL_2017} S. Kunisada, S. Adachi, S. Sakai, N. Sasaki, M. Nakayama, S. Akebi, K. Kuroda, T. Sasagawa, T. Watanabe, S. Shin, and T. Kondo, Phys. Rev. Lett. {\bf 119}, 217001 (2017).

\bibitem{Ding_EPL_2008} H. Ding, P. Richard, K. Nakayama, K. Sugawara, T. Arakane, Y. Sekiba, A. Takayama, S. Souma, T. Sato, and T. Takahashi, Europhys. Lett. {\bf 83} 47001 (2008).

\bibitem{Evtushinsky_NJP_2009}  D. V. Evtushinsky, D. S. Inosov, V. B. Zabolotnyy, M. S. Viazovska, R. Khasanov, A. Amato, H. -H. Klauss, H. Luetkens, Ch. Niedermayer, G. L. Sun, V. Hinkov, C. T. Lin, A. Varykhalov, A. Koitzsch, M. Knupfer, B. Büchner, A. A. Kordyuk, and S. V. Borisenko, New J. Phys. {\bf 11}, 055069 (2009).

\bibitem{Khasanov_Ba122_PRL_2009} R. Khasanov, D. V. Evtushinsky, A. Amato, H. -H. Klauss, H. Luetkens, Ch. Niedermayer, B. Büchner, G. L. Sun, C. T. Lin, J. T. Park, D. S. Inosov, and V. Hinkov, Phys. Rev. Lett. {\bf 102}, 187005 (2009).


\bibitem{Khasanov_Sr122_PRL_2009} R. Khasanov, A. Maisuradze, H. Maeter, A. Kwadrin, H. Luetkens, A. Amato, W. Schnelle, H. Rosner, A. Leithe-Jasper, and H.-H. Klauss, Phys. Rev. Lett. {\bf 103}, 067010 (2009).

\bibitem{Khasanov_SrPtP_PRB_2014} R. Khasanov, A. Amato, P. K. Biswas, H. Luetkens, N. D. Zhigadlo, and B. Batlogg, Phys. Rev. B {\bf 90}, 140507(R) (2014).

\bibitem{Khasanov_1144_PRB_2019} R. Khasanov, W. R. Meier, S. L. Bud'ko, H. Luetkens, P. C. Canfield, and A.  Amato, Phys. Rev. B {\bf 99}, 140507(R) (2019.)


\bibitem{Ruby_PRL_2015} M. Ruby, B.W. Heinrich, J.I. Pascual, and K.J. Franke, Phys. Rev. Lett. {\bf 114}, 157001 (2015).

\bibitem{Singh_PRB_2018} J. Singh, A. Jayaraj, D. Srivastava, S. Gayen, A. Thamizhavel, and Y. Singh, Phys. Rev. B {\bf 97}, 054506 (2018).



\bibitem{Amato_GPS_2017} A. Amato, H. Luetkens, K. Sedlak, A. Stoykov, R. Scheuermann, M. Elender, A. Raselli, D. Graf, Rev. Sci. Instrum. {\bf 88}, 093301 (2017).

\bibitem{Tinkham_book_1975} M. Tinkham, {\it Introduction to Superconductivity} (Krieger Publishing company, Malabar, Florida, 1975).

\bibitem{deGennes_Book_1966} P.G. de Gennes, {\it Superconductivity of Metals and Alloys} (Benjamin, New-York, 1966).

\bibitem{Kittel_Book_1996} C. Kittel, {\it Introduction to Solid State Physics, 7th Ed.}, (Wiley, India, Pvt. Limited, 2007).

\bibitem{Prozorov_PRL_2007} R. Prozorov, Phys. Rev. Lett. {\bf 98}, 257001 (2007).

\bibitem{Prozorov_NatPhys_2008} R. Prozorov, A. F. Fidler, J. R. Hoberg, and P. C. Canfield, Nature Phys. {\bf 4}, 327 (2008).

\bibitem{Khasanov_Bi-II_PRB_2019} R. Khasanov, M. M. Radonji\'{c}, H. Luetkens, E. Morenzoni, G. Simutis, S. Sch\"{o}necker, W. H. Appelt, A. \"{O}stlin, L. Chioncel, and A. Amato, Phys. Rev. B {\bf 99}, 174506 (2019).

\bibitem{Karl_PRB_2019} R. Karl, F. Burri, A. Amato, M. Doneg\`{a}, S. Gvasaliya, H. Luetkens, E. Morenzoni, and R. Khasanov, Phys. Rev. B {\bf 99}, 184515 (2019).

\bibitem{Khasanov_Ga-II_arxiv} R. Khasanov, H. Luetkens, A. Amato, and E. Morenzoni, Phys. Rev. B {\bf 101}, 054504 (2020).

\bibitem{Supplemental_part} The Supplementary part of the manuscript describes the TF-$\mu$SR data analysis procedure, and the results of the fit of the single gap $\alpha-$model to the experimental $B_{\rm c}(T)$ data. The Supplemental part includes References \onlinecite{Suter_MUSRFIT_2012,Prozorov_PRAppl_2018,Yaouanc_book_2011,Carrington_2003}.

\bibitem{Padamsee_JLTP_1973} H. Padamsee, J. E. Neighbor, and C. A. Shiffman, J. Low Temp. Phys. {\bf 12}, 387 (1973).

\bibitem{Johnston_SST_2013} D.C. Johnston,  Supercond. Sci. Technol. {\bf 26}, 115011 (2013).

\bibitem{MathCad} https://www.mathcad.com/

\bibitem{Muehlschlegel_ZPhys_1959} B. M\"uhlschlegel, Z. Phys. {\bf 155}, 313, {\bf 1959}.

\bibitem{Bussmann-Holder_EPB_2004} A. Bussmann-Holder, R. Micnas, and A. R. Bishop, Eur. Phys. J. B. {\bf 37}, 345 (2004).

\bibitem{Bussmann-Holder_Arxiv_2009} A. Bussmann-Holder, arXiv:0909.3603, unpublished.

\bibitem{Khasanov_PRL_2010} R. Khasanov, M. Bendele, A. Amato, K. Conder, H. Keller, H.-H. Klauss, H. Luetkens, and E. Pomjakushina, Phys. Rev. Lett. 104, 087004 (2010).

\bibitem{Kogan_PRB_2009} V. G. Kogan, C. Martin, and R. Prozorov, Phys. Rev. B {\bf 80}, 014507 (2009).

\bibitem{Gupta_arxiv_2019} R. Gupta, A. Maisuradze, N.D. Zhigadlo, H. Luetkens, A. Amato, and R. Khasanov, Front. Phys. {\bf 8}, 2 (2020).

\bibitem{Kim_Symmetry_2019} H. Kim, K. Cho, M.A. Tanatar, V. Taufour, S.K. Kim, S.L. Bud’ko, P.C. Canfield, V.G. Kogan, and R. Prozorov Symmetry {\bf 11}, 1012 (2019).

\bibitem{Bang_PRB_2009} Y. Bang, H.-Y. Choi, and H. Won, Phys. Rev. B {\bf 79}, 054529 (2009).

\bibitem{Trevisan_PRB_2018} T.V. Trevisan, M. Sch\"{u}tt, and R.M. Fernandes, Phys. Rev. B {\bf 98}, 094514 (2018).




\bibitem{Suter_MUSRFIT_2012} A. Suter and B. M. Wojek, Phys. Procedia {\bf 30}, 69 (2012).


\bibitem{Prozorov_PRAppl_2018} R. Prozorov and V.G. Kogan, Phys. Rev. Applied {\bf 10}, 014030 (2018).

\bibitem{Yaouanc_book_2011} A. Yaouanc, and P. Dalmas de R\'{e}otier, {\it Muon Spin Rotation, Relaxation and Resonance: Applications to Condensed Matter} (Oxford University Press, Oxford, 2011).

\bibitem{Carrington_2003} A.~Carrington and F.~Manzano, Physica~C {\bf 385}, 205 (2003).




\end{thebibliography}
\end{document}